\documentclass[aps,prb,showpacs,twocolumn,superscriptaddress]{revtex4-2}

\usepackage{bm,here,color}
\usepackage{graphicx}
\usepackage{amsmath,amssymb}
\usepackage{braket}
\usepackage{siunitx}
\usepackage{hyperref}
\begin{document}
\title{Predicted versatile topological nodal magnons \\ in Tb-based icosahedral quasicrystal 1/1 approximants}
\author{Rintaro Eto}
\email{rintaro.wufse@toki.waseda.jp}
\affiliation{Department of Applied Physics, Waseda University, Okubo, Shinjuku-ku, Tokyo 169-8555, Japan}
\author{Masahito Mochizuki}
\email{masa_mochizuki@waseda.jp}
\affiliation{Department of Applied Physics, Waseda University, Okubo, Shinjuku-ku, Tokyo 169-8555, Japan}
\author{Shinji Watanabe}
\email{swata@mns.kyutech.ac.jp}
\affiliation{Department of Basic Sciences, Kyushu Institute of Technology, Kitakyushu, Fukuoka 804-8550, Japan}
\date{\today} 
\begin{abstract}
    Using a recently-established band representation analysis, we discover two distinct types of topological nodal magnons in the real-space antiferroic ordering of whirling spin arrangements in the Tb-based icosahedral quasicrystal 1/1 approximants, both of which originate from a composite band (co-)representation $A\uparrow P_In\bar{3}(24)$ and its constituent elementary band representations. The first type is doubly-degenerate nodal line network and nodal planes associated with two-dimensional irreducible band representation, while the second type is a nodal line network due to accidental band inversions. Since our analysis, which relies solely on magnetocrystalline symmetry, is valid for a wide range of materials and spin textures belonging to the same magnetic space group irrespective of composition, these findings offer new universal insights into the research of Tb-based quasicrystal approximants as well as a contribution to broadening the range of topological magnon-hosting materials.
\end{abstract}
\maketitle
\section{Introduction}
\begin{figure}[htb]
    \centering
    \includegraphics[scale=1.0]{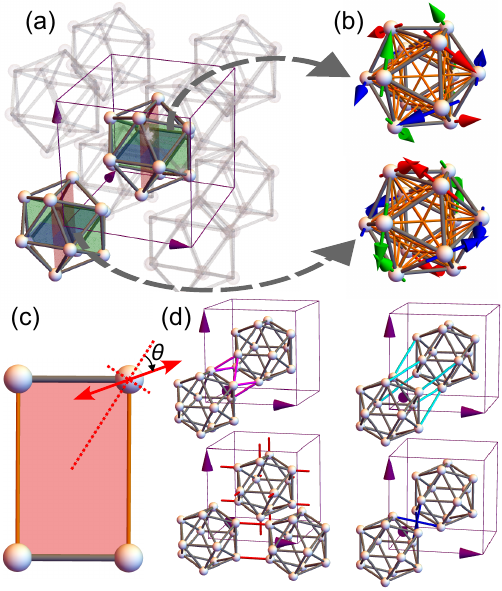}
    \caption{(a) Lattice structure of the icosahedral quasicrystal 1/1 approximant. The cube indicates the magnetic unit cell for antiferroic orders. Each icosahedron at bcc lattice points can be decomposed into three (red, green, and blue) rectangles with their aspect ratio of $\tau=\left(1+\sqrt{5}\right)/2$. (b) Icosahedral clusters with whirling and antiwhirling spin arrangements, where $\theta=\ang{86}$. Gray and orange bonds are intracluster nearest-neighbor and next-nearest-neighbor bonds $J_1$ and $J_2$, respectively. Red, green, and blue arrows represent the localized spins of the rare-earth atoms placed at the vertices of red, green, and blue rectangles in (a), respectively. (c) Definition of the angle of the monoaxial anisotropy $\theta$, where the red rectangle corresponds to the one in (a). (d) Visualized intercluster bonds $J_{1a}$ (magenta), $J_{2a}$ (cyan), $J_{1b}$ (red), and $J_{2b}$ (blue).}
    \label{Fig01}
\end{figure}
The topological electronic states in materials~\cite{Hasan2010,QiXL2011,Bernevig2022} have long been the subject of intensive research, primarily due to their novel functionality, which is applicable to energy-efficient computational devices and information carriers. Recently, their magnetic counterpart, referred to as the ``topological magnon"~\cite{McClarty2022,Karaki2023,Reichlova2024}, has also become a fascinating research topic due to potential applications in spintronics. Various magnonic counterparts of topological electoronic states, such as Chern insulators~\cite{ZhangL2013,Mook2014Kagome,Mook2014EdgeState}, $\mathbb{Z}_2$ topological insulators~\cite{Kondo2019SH,Kondo2019_3D}, Weyl nodes~\cite{LiFY2016,Mook2016,ZhangLC2020}, nodal lines~\cite{Mook2017,Pershoguba2018,YuanB2020,Scheie2022}, and topological crystalline insulators~\cite{Hirosawa2020,Kondo2021}, have been revealed.
\par Quasicrystal (QC), which is a class of arrangements of atoms only with long-range orientation but without translational symmetry, has been intensively studied since its first discovery in the Al-Mn alloy by Shechtman {\it et al}.~\cite{Shechtman1984,Shechtman1985} The lack of periodicity in QCs gives rise to unique electronic states, which have continued to be fascinating research topics for more than 40 years since their discovery. Moreover, QCs have phason distortion~\cite{Socolar1986,Yamamoto1996} as well as phonon distortion, which could offer a platform of a variety of anomalous thermodynamic effects~\cite{Huang2024,Nagai2024}, structural transitions~\cite{Chattopadhyay1987,Ishii1989,ZhangH1990}, and lattice dynamics~\cite{Boissieu2007,Matsuura2024}.
\par Magnetic properties in QCs are also fascinating research topics, and a long-standing question remains unresolved for a few decades: Can the magnetic long-range order appear in QCs? In the quest for its experimental identification, numerous trials have been dedicated not only to QCs but also to their approximant crystals (ACs)~\cite{Goldman1993} with translational symmetry in a rigorous sense~\cite{Suzuki2021R}. These ACs and QCs share the same concentric shell structures of atomic polyhedrons known as Tsai-type clusters~\cite{Tsai2000}. Inside these Tsai-type clusters, rare-earth atoms, which possess magnetic moments, are located at each vertex of the icosahedron shown in Figs.~\ref{Fig01}(a) and \ref{Fig01}(b). Antiferromagnetic order has been observed in the 1/1 AC Cd$_6R$ ($R=$Tb, Y, Pr, Nd, Sm, Gd, Tb, Dy, Ho, Er, Tm, Yb, and Lu)~\cite{Tamura2010,Mori2012} and in the 1/1 AC Au-Al-$R$ ($R=$ Gd and Tb)~\cite{Ishikawa2018}. Ferromagnetic order has been observed not only in 1/1 ACs~\cite{Hiroto2013,Hiroto2014,Hiroto2020} but also in QCs~\cite{Tamura2021,Takeuchi2023}. However, clarifying the stabilization mechanism of these magnetic orderings remains challenging, and there is still no consensus. Theoretical analyses have been limited to those with extremely simple spin models involving isotropic short-range and/or long-range Ruderman-Kittel-Kasuya-Yosida (RKKY) type interactions~\cite{Axenovich2001,Konstantinidis2005,Miyazaki2020,Suzuki2021}.
\par Under these circumstances, one of the authors (S.W.) recently performed a theoretical analysis based on the point-charge model~\cite{Watanabe2021,Watanabe2021PNAS}. This approach incorporates the crystalline electric field (CEF) effects at the rare-earth sites, enabling the construction of realistic spin models that account for anisotropy in spin space. In fact, the derived spin model for 1/1 ACs describes the stabilization of a ferroic order of the ``hedgehog" spin textures which exhibit nontrivial real-space topological charges of $\pm1$, as well as the whirling-antiwhirling order, which has a large topological charge of $\pm 3$ and has been experimentally observed in Au$_{72}$Al$_{14}$Tb$_{14}$ by T. J. Sato {\it et al}. recently~\cite{Sato2019}. Additionally, the derived spin model is valuable for studying magnetic excitation properties. Indeed, emergence of nonreciprocal magnon bands in the aforementioned ferroic hedgehog state without inversion symmetry in 1/1 ACs is clarified~\cite{Watanabe2023}.
\par In this paper, we theoretically reveal emergence of topological magnon excitations in 1/1 ACs by analyzing this realistic spin model with the band representation theory~\cite{Slager2013,Kruthoff2017}. Specifically, we investigate the magnetic excitations in the whirling-antiwhirling magnetic order of the rare-earth-based icosahedral quasicrystal 1/1 ACs by using the linear spin-wave theory (LSW) for this simple but realistic spin model and symmetry analysis based on the magnetic topological quantum chemistry (MTQC) database on the Bilbao Crystallographic Server~\cite{Bradlyn2017,Vergniory2017,XuY2020,Elcoro2021,Cano2021,Corticelli2023}. Our analysis clarifies versatile topological nodal lines and planes of magnons over the entire cubic Brillouin zone. Some of them are protected purely by magnetocrystalline symmetry, and others are induced by accidental band inversions under partial supports by symmetry. This symmetry analysis gives a universal result for a specific magnetic space group and a Wyckoff position (WP), and thus, it is applicable to different compositions. This work indicates that 1/1 ACs, which have vast diversity of materials, are promising candidates for topological magnon-hosting materials, and thus, are useful for spintronics.
\section{Model}
We start with the simple spin model for the icosahedral quasicrystal 1/1 ACs, which is derived from the point-charge model analysis taking the effect of CEFs of ligand ions surrounding the rare-earth atoms into account~\cite{Watanabe2021PNAS}. Note that, in the original point-charge model, we neglect orbital anisotropy of ligand ions, and thus their impact on the exchange interactions is not taken into account. We also neglect anisotropy coming from the intercluster structures~\cite{Thiem2015a,Thiem2015b,Jeon2025}. However, these effects do not disturb our arguments relying solely on magnetocrystalline symmetry presented in the remainder of this paper. Here we take Au-Si-Tb or Au-Al-Tb as target materials just to be described by this model. The Hamiltonian is given by
\begin{equation}
    \mathcal{H}_{\rm spin}(\theta) = \sum_{{\bm r},{\bm r}'} J_{{\bm r},{\bm r}'}{\bm S}_{{\bm r}}\cdot{\bm S}_{{\bm r}'} - D\sum_{\bm r}\left({\bm S}_{{\bm r}}\cdot{\bm e}_{{\bm r}}(\theta)\right)^2
    \label{eq:HmSpin}
\end{equation}
where ${\bm S}_{\bm r}$ denotes the localized spin on the rare-earth site located at vertices of icosahedrons ${\bm r}$. The first and second terms describe the exchange coupling between spins at ${\bm r}$ and ${\bm r}'$ and the single-ion monoaxial anisotropy along the site-dependent unit vector ${\bm e}_{\bm r}(\theta)$, respectively. The definition of the anisotropy angle $\theta$ is shown in Fig.~\ref{Fig01}(c). The monoaxial anisotropy comes from the CEFs of ligand ions, and its magnitude $D$ is typically much larger than the exchange couplings $J_{{\bm r},{\bm r}'}$ $\left(D\gg|J_{{\bm r},{\bm r}'}|\right)$ in Tb-based materials. Note that the anisotropy angle $\theta$ is determined by the ratio $Z_{\rm Si(Al)}/Z_{\rm Au}$ where $Z_{\rm Si(Al)}$ and $Z_{\rm Au}$ are the valences of Si (Al) and Au atoms, respectively.
Because of this dominant monoaxial anisotropy, it is assumed that the ground-state spin arrangement is always parallel or antiparallel to the anisotropy axis, i.e., ${\bm S}_{\bm r}\parallel\pm{\bm e}_{\bm r}(\theta)$. \par
It is also worth noting that, in general, multipolar fluctuations~\cite{Jeon2024} can play crucial roles in icosahedral clusters, which potentially undermine the validity of our simple spin model given by Eq.~(\ref{eq:HmSpin}). However, dominant monoaxial anisotropy in Tb atoms essentially stabilize magnetic order with finite local dipolar moment on each Tb atom, and suppress higher-rank multipolar fluctuations. In this sense, we believe that our simple spin model and noninteracting LSW theory introduced in the following are valid, at least, for Tb-based materials. \par
In the previous work by one of the authors (S.W.), the ground-state spin arrangement of the model in Eq.~(\ref{eq:HmSpin}) under the assumption of ${\bm S}_{\bm r}\parallel\pm{\bm e}_{\bm r}(\theta)$ has been revealed by comparing the energies of all possible spin configurations~\cite{Watanabe_submitted}. When we consider the exchange coupling of $J_2=J_{2a}=J_{2b}=\alpha J_1=\alpha J_{1a}=\alpha J_{1b}$ with $\alpha\gtrsim10$ and an anisotropy angle $\theta\sim90^\circ$, the ground-state spin texture is the whirling-antiwhirling state. 
This is consistent with the experimental evaluation of $J_2/J_1\gg 1$ for the whirling-antiwhirling state observed in the 1/1 AC Au-Ga-Tb, which was estimated from the fitting of the measured temperature and field dependences of the magnetization on the basis of the effective model on {\it single} icosahedron~\cite{Labib2024,Nawa2023}. Namely, the monoaxial effective model on simplified {\it single} icosahedron supports the above numerically exact result in the 1/1 AC where the whirling-antiwhirling state is realized for $J_2/J_1\gg 1$.
Note that in this whirling-antiwhirling (antiferromagnetic) state, the spin arrangements on the icosahedral clusters located at the corners of the cubic magnetic unit cell and those located at the center of the unit cell are perfectly opposite to each other. These arrangements are connected by a combined operation of $\mathcal{T}{\bm \tau}$, where $\mathcal{T}$ represents the time-reversal operation and ${\bm \tau}$ is the incomplete translation that connects the cluster at the cell corner to that at the cell center [See Fig.~\ref{Fig01}(a)]. This spin arrangement corresponds to the type-IV magnetic space group (MSG) $P_In\bar{3}$, which is \#201.21 in the Belov-Neronova-Smirnova (BNS) notation. We remark that this magnetic space group is also denoted as $I_{\rm P}m'\bar{3'}$ (No. 204.5.1534 in the Opechowski-Guccione (OG) notation)~\cite{Watanabe_submitted}. 
\par
To obtain the magnon band dispersion within the harmonic level, we apply the linear spin-wave (LSW) theory to this model. First, we bosonize the spin Hamiltonian $\mathcal{H}_{\rm spin}(\theta)$ by using the truncated Holstein-Primakoff transformation: ${\bm S}_{\bm r}\rightarrow\hat{\mathsf{S}}_{\bm r}^+{\bm e}_{\bm r}^- + \hat{\mathsf{S}}_{\bm r}^-{\bm e}_{\bm r}^+ + \hat{\mathsf{S}}_{\bm r}^0{\bm e}_{\bm r}^0$ with
\begin{equation}
    \hat{\mathsf{S}}^+_{\bm r} \sim \sqrt{S}\hat{a}_{\bm r}, \quad \hat{\mathsf{S}}^-_{\bm r} \sim \sqrt{S}\hat{a}_{\bm r}^\dagger, \quad \hat{\mathsf{S}}^0_{\bm r} \sim S-\hat{a}^\dagger_{\bm r}\hat{a}_{\bm r},
\end{equation}
where $\hat{a}_{\bm r}$ $\left(\hat{a}^\dagger_{\bm r}\right)$ corresponds to the magnon annihilation (creation) operator on the site ${\bm r}$. Note that ${\bm e}^\pm_{\bm r}\equiv\left({\bm e}^1_{\bm r}\pm i{\bm e}^2_{\bm r}\right)/\sqrt{2}$ and ${\bm e}^0_{\bm r}\left(\parallel{\bm S}_{\bm r}\right)$ are complex and real unit vectors respectively, where ${\bm e}^1_{\bm r}\times{\bm e}^2_{\bm r}={\bm e}^0_{\bm r}$ is satisfied. Then we obtain the quadratic Hamiltonian in terms of magnon operators $\hat{a}_{\bm r}$ and $\hat{a}^\dagger_{\bm r}$. Finally, through diagonalization (Bogoliubov transformation) of this quadratic Hamiltonian using the procedure proposed by Colpa~\cite{Colpa1978}, we obtain the magnon band dispersion relations within the harmonic approximation. 
Note that large monoaxial anisotropy $D$ validates the harmonic approximation.
In addition, since magnon-magnon interactions cannot lift magnon band degeneracies exhibited by noncoplanar spin textures~\cite{Gohlke2023}, the harmonic LSW theory is accurate enough to capture nodal degeneracy physics qualitatively.
\section{MTQC analysis}
\begin{figure}[htb]
    \centering
    \includegraphics[scale=1.0]{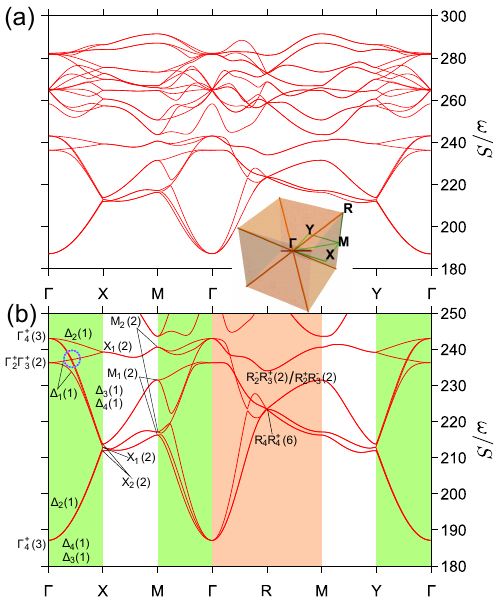}
    \caption{(a) Magnon band dispersion in the spin model given by Eq.~(\ref{eq:HmSpin}). Parameter set used is presented in the main text. Inset shows the magnetic Brillouin zone. Orange lines and planes correspond to the type-A nodal lines and planes, respectively. Green lines show the momentum path used in both (a) and (b). (b) Enlarged view of the lower-lying eight bands. Irreps of the bands are also given. Orange backgrounds indicate the paths with type-A nodal structures, while green ones indicate the paths on the glide planes.}
    \label{Fig02}
\end{figure}
\begin{figure*}[htb]
    \centering
    \includegraphics[scale=1.0]{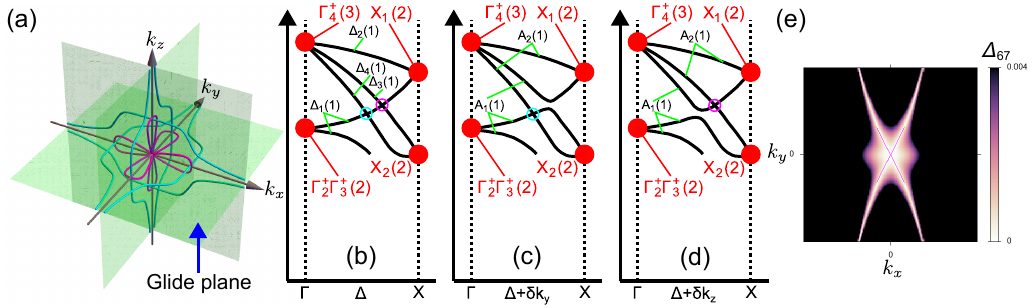}
    \caption{(a) Magenta and cyan lines are trajectories of nodal lines in the lower sector $^1E_g\uparrow G$ $\left({\rm or} \ ^2E_g\uparrow G(8) \right)$, which are also highlighted by circles in (b-1)-(b-3) and Fig.~\ref{Fig02}(a). (b)-(d) Schematics of band dispersions (b) along the path $\Gamma$-X referred to as $\Delta$, (c) the path slightly deviated from $\Gamma$ towards $k_y$-direction (with the starting and ending points corresponding to the $\Gamma$ and X points, respectively), and (d) the path slightly deviated from $\Gamma$ towards $k_z$-direction. Symmetry eigenvalues of irreps on the path $\Delta$ are summarized in the Supplemental Material Table~SII~\cite{SuppMat}. (e) Bandgap between the sixth and seventh bands. White regions indicate the gapless areas and correspond to the magenta nodal lines in (a) that cross at $\Gamma$. Magenta lines indicate nodal-line solutions with $k_z=0$ derived by diagonalizing the low-energy effective Hamiltonian $\mathcal{H}^{\rm eff}_{\Gamma_4^\pm(3)}({\bm k})$ given by Eq.~(\ref{eq:LowEnergyEffHm}).}
    \label{Fig03}
\end{figure*}
We first search for the single-valued band representation for the non-maximal WP $24g$ of the MSG $P_In\bar{3}$ (BNS \#201.21)~\cite{Corticelli2023}. Band irreps constituting this band representation are summarized in the Supplementary Material (SM) Table~SI~\cite{SuppMat}. This is essentially the magnon band representation of the whirling-antiwhirling state in the 1/1 AC. Referring to the MTQC database on the Bilbao Crystallographic Server, we find the only band representation satisfying all these conditions is a composite representation $A\uparrow G(24)$, where $G=P_In\bar{3}$. It is also worth noting that $2A$ is the transverse orbital irrep of the unitary subgroup 1 of the magnetic point group $m'$, which is isomorphic to the magnetic site-symmetry group of the WP $24g$~\cite{Corticelli2023}. The factor 2 arises from the redundancy of the paraunitary nature of the Bogoliubov transformation. \par
We present the typical magnon band dispersion in the whirling-antiwhirling state in Fig.~\ref{Fig02}(a). Hereafter we set $\theta=86^\circ$ which corresponds to the experimentally observed whirling-antiwhirling order in Au${}_{72}$Al${}_{14}$Tb${}_{14}$. Parameter values of the exchange couplings $J_1,J_2,J_{1a},J_{1b},J_{2a}$, and $J_{2b}$ shown in Figs.~\ref{Fig01}(a), \ref{Fig01}(b), and \ref{Fig01}(d), and the site-dependent monoaxial anisotropy $D$ are given as $J_1=-1$, $J_2=-10$, $D=100$, $J_{1a}=0.98J_1$, $J_{1b}=0.96J_1$, $J_{2a}=0.98J_2$, and $J_{2b}=0.96J_2$, respectively. Note that this parameter set assumes highly simplified case $J_1=J_{1a}=J_{1b}$ and $J_2=J_{2a}=J_{2b}$, where the exchange parameters depend only on the distance of neighboring atoms (Ruderman-Kittel-Kasuya-Yosida-like). We applied tiny modification to this simplified case just to avoid unphysical spurious degeneracies. We confirmed that smaller $D$ does not change the entire band dispersion so much [See the SM Fig.~S1(a)~\cite{SuppMat}].  Because there are two icosahedron clusters and hence 24 rare-earth atoms in a magnetic unit cell, the excitation spectrum is composed of 24 bands. The composite representation $A\uparrow G(24)$ gives specific sets of irreps for every crystal moment, for example, on the $\Gamma$-R path along [111]:
\begin{equation}
    \Lambda:(u,u,u) = 8\Lambda_1(1) \oplus 8\Lambda_2\Lambda_3(2).
    \label{eq:IrrepGR}
\end{equation}
Equation~(\ref{eq:IrrepGR}) indicates that there are eight non-degenerate bands and eight doubly-degenerate nodal lines on the $\Gamma$-R path. In addition, we can find 12 doubly-degenerate nodal planes on the edges of the magnetic Brillouin zone, whose representation is given by
\begin{equation}
    S:(u,1/2,v) = 12S_1S_2(2).
    \label{eq:IrrepNodalPlane}
\end{equation}
The arrangement of the nodal lines and planes given by Eqs.~(\ref{eq:IrrepGR}) and (\ref{eq:IrrepNodalPlane}) are shown in the inset of Fig.~\ref{Fig02}(a). It is important that all these nodal structures originate from high (more than one)-dimensional irreps, which we refer to as type-A nodes in the following. \par
The representation of the lower-lying eight bands in Fig.~\ref{Fig02}(a), which are separated from the others by a global bandgap, is equivalent to an indecomposable elementary band representation $^1E_g\uparrow G(8)$ $\left({\rm or} \ ^2E_g\uparrow G(8) \right)$, where $^1E_g\oplus^2E_g$ is the transverse orbital irrep of the unitary subgroup $\bar{3}$ of the magnetic point group $\bar{3}$, which is isomorphic to the magnetic site-symmetry group of the maximal WP $8c$. Since the WP $8c$ is not compatible with any magnetically ordered states belonging to the MSG $P_In\bar{3}$, this is one of the nontrivial realizations of the magnon band representation induced by the incompatible WP $8c$. In the following, we focus on this lower sector with a representation $^1E_g\uparrow G(8)$ $\left({\rm or} \ ^2E_g\uparrow G(8) \right)$. Figure~\ref{Fig02}(b) provides an enlarged view for the lower sector, where irreps of the magnon bands are indicated. \par
By examining the compatibility relations, accidental band inversions at high-symmetry momenta, and band irreps themselves, we identify additional nodal structures which are fundamentally different from those of type-A. We focus specially on the two nodal lines indicated by a circle in Fig.~\ref{Fig02}(b). Spatial evolution of these nodal lines in the Brillouin zone are shown in Fig.~\ref{Fig03}(a). These nodal lines are associated with an accidental band inversion between X$_1$(2) and X$_2$(2) presented in Figs.~\ref{Fig03}(b)-(d) and are protected by glide symmetries. Because of this protection by glides, the nodal lines reside within the $k_x$-, $k_y$-, and $k_z$-planes, each of which is a glide plane. We hereafter refer to them as type-B nodal lines.
\par From symmetry consideration, we derive a low-energy effective Hamiltonian for the three-dimensional representation at the $\Gamma$ point $\Gamma_4^\pm(3)$, which is given in the form,
\begin{equation}
    \mathcal{H}^{\rm eff}_{\Gamma_4^\pm(3)}({\bm k}) = \left(
        \begin{array}{ccc}
            \mathcal{F}_1({\bm k}) & D'k_xk_y & D'k_zk_x \\
            D'k_xk_y & \mathcal{F}_2({\bm k}) & D'k_yk_z \\
            D'k_zk_x & D'k_yk_z & \mathcal{F}_3({\bm k})
        \end{array}
    \right),
    \label{eq:LowEnergyEffHm}
\end{equation}
where $\mathcal{F}_1({\bm k})=Ak_x^2+Bk_y^2+Ck_z^2$, $\mathcal{F}_2({\bm k})=Ck_x^2+Ak_y^2+Bk_z^2$, $\mathcal{F}_3({\bm k})=Bk_x^2+Ck_y^2+Ak_z^2$, and $A,B,C,D'\in\mathbb{R}$. Note that the Hamiltonian is irrelevant to the parity $\pm$.
\par This effective Hamiltonian does describe the emergence of some of the above-discussed nodal lines. First, we obtain eigenenergies along [111] ($\Gamma$-R path $\Lambda$, $k_x=k_y=k_z=k$), which is given by
\begin{equation}
\begin{aligned}
    E &= \underbrace{(A+B+C-D')k^2}_{\Lambda_2\Lambda_3(2)}, \quad \underbrace{(A+B+C+2D')k^2}_{\Lambda_1(1)}.
\end{aligned}
\end{equation}
The former is doubly-degenerate solution and corresponds to a type-A nodal line. The set of these solutions satisfies the compatibility relation $\Gamma\leftrightarrow\Lambda$. Next, we obtain eigenenergies on the glide plane ($k_z=0$), which is given by
\begin{equation}
\begin{aligned}
    E &= Bk_x^2+Ck_y^2, \\
    &\quad\quad \frac{1}{2} \left\{ (A+C)k_x^2+(A+B)k_y^2 \right\} \\
    &\quad\quad \ \pm\frac{1}{2} \sqrt{ \left\{ (A+C)k_x^2+(B-A)k_y^2 \right\}^2+4D'^2k_x^2k_y^2 }.
\end{aligned}
\end{equation}
Within the long-wavelength limit $(k_x,k_y)=(k,\beta k)$ $(\beta\in\mathbb{R})$, the nodal-line solution is given by
\begin{equation}
    \beta = \pm \sqrt{\frac{-F_a\pm\sqrt{F_a^2-4F_bF_c}}{2(A-C)(B-C)}}
    \label{eq:effHmSol_beta}
\end{equation}
where $F_a=A^2-B^2-C^2-AB-AC+3BC-D'^2$, $F_b=AC-AB-BC+B^2$, and $F_c=AB-AC-BC+C^2$. By numerical fitting for the upper $\Gamma_4^+(3)$ bands in Fig.~\ref{Fig02}(b), we obtain $A\sim-0.575$, $B\sim-2.219$, $C\sim-2.374$ and $D'\sim1.995$, which yields $\beta\sim\pm1.996$. As seen in Fig.~\ref{Fig03}(e), this long-wavelength solution accurately fits type-B nodal lines indicated by magenta lines in Fig.~\ref{Fig03}(a). Note that, in general, this crossing nodal lines appear only when the solutions $\beta$ in Eq.~(\ref{eq:effHmSol_beta}) are real. Interestingly, we found that real solutions of $\beta$ can be obtained from more than 87\% region of the parameter space hypercube $\{A,B,C,D'\}\in[-v_\mathrm{lim},v_\mathrm{lim}]^4$ $(v_\mathrm{lim}>0)$, which indicates that accidental band inversions to realize type-B nodal degeneracies occur very frequently in the model parameter space.
It is also worth noting that, by applying different model parameters $J_{{\bm r},{\bm r}'}$ and $D$ than those for Fig.~\ref{Fig02}, another EBR, $B_1\uparrow G(6)$ (or $B_2\uparrow G(6)$), appears as a lower-lying sector. We present band dispersion with the lower-lying sector $B_1\uparrow G(6)$ (or $B_2\uparrow G(6)$) in the SM Fig.~S1(b)~\cite{SuppMat}. Even in this case, both symmetry-enforced type-A and accidental type-B nodal degeneracies appear. This analysis also supports our claim that emergence of accidental type-B nodal lines, which are not enforced by symmetry but merely accidental, seems to happen frequently.
\par Taken together, the three-dimensional representation $\Gamma_4^\pm(3)$ are closely related with both type-A and -B nodal lines. This aspect highlights the importance of detailed magnetocrystalline symmetry properties for the emergence of topological magnons in quasicrystal 1/1 ACs. Since the long-wavelength effective theory describes these nodal lines accurately, long-wavelength measurements would be useful for their experimental identification. In addition, both type-A and type-B nodal lines are characterized by the same topological invariant, the quantized Berry phase of $\pi$~\cite{Mook2017}. This indicates intensity winding of dynamical spin structure factor, which can be observed by neutron scattering~\cite{Scheie2022}, would be also useful for experimental detection of our predicted nodal degeneracies. \par
\section{Summary}
In summary, we have theoretically clarified possible manifestation of topological magnon excitations in the whirling-antiwhirling magnetic state in the icosahedral quasicrystal 1/1 ACs. On the basis of the LSW+MTQC analysis, we have found two distinct types of topological nodal magnons referred to as type-A and type-B. It has been revealed that the former originates purely from magnetocrystalline symmetry, while the latter requires an accidental band inversion for their emergence. 
We have also clarified that the long-wavelength effective theory accurately captures the two distinct types of topological nodal magnons, which indicates that long-wavelength measurements are promising for investigating a variety of nodal line physics in ACs.
\par Finally, we note that the exploration of materials for quasicrystal approximants with magnetism is rapidly advanced. Even merely with combinations of elements, a vast number of possibilities can be explored. Moreover, their composition ratios can be continuously varied for each. Thus, it can be said that the number of achievable compounds is literally diverse. In this sense, this work, which reveals the possible emergence of topological magnons in ACs for the first time, paves the new pathway to research in the field of magnetism in ACs.
\section{Data availability}
Codes and data are available from the corresponding author (R.E.) upon reasonable request.
\section{Acknowledgment}
This work was supported by JSPS KAKENHI (Grants No. 20H00337, No. 22H04597, No. 22H01170, No. 23H04522, No. 23K17672, and No. 24H02231) and JST CREST (Grant No. JPMJCR20T1). M.M. was supported by Waseda University Grant for Special Research Projects (Grants No. 2023C-140 and No. 2024C-153). R.E. was supported by a Grant-in-Aid for JSPS Fellows (Grant No. 23KJ2047). A part of the numerical simulations was carried out at the Supercomputer Center, Institute for Solid State Physics, University of Tokyo. \par

\end{document}